\documentclass[published]{JHEP}
\JHEP{01(1999)005}
 
\def\beq{\begin{equation}}
\def\eeq{\end{equation}}
\def\bea{\begin{eqnarray}}
\def\eea{\end{eqnarray}}
\def\bq{\begin{quote}}
\def\eq{\end{quote}}

\def\Det{{\rm Det}}

\title{Four-dimensional QCD and fiberwise duality}

\author{Marco Bochicchio\\
	INFN Sezione di Roma, Dipartimento di Fisica,\\
	 Universit\`a di Roma ``La Sapienza'', \\
	Piazzale Aldo Moro 2 , 00185 Roma  \\ 
	E-mail: \email{Marco.Bochicchio@roma1.infn.it}}
\abstract{
We transform, by means of a fiberwise duality, the partition function of QCD
on a product of two two-tori, into a four-dimensional $\sigma$-model, whose 
target space is the cotangent space of unitary connections on the fiber torus 
fiberwise.}
\received{October 9, 1998}
\accepted{January 8, 1998}
 
\begin{document}
          
\renewcommand{\baselinestretch}{1.06}\normalsize
\section{Introduction}

The prominent role that integrable systems (Hitchin systems~\cite{H}), defined
over the cotangent bundle of unitary connections on a Riemann surface, play in
the solution~\cite{SW} of the low energy limit of four-dimensional
${\cal N}=2$
supersymmetric gauge theories and certain string 
theories~\cite{D}, has stimulated efforts to extend to other non-supersymmetric
theories the $ {\cal N}=2$ techniques~\cite{A}.

Yet, the field theoretical origin, in the four-dimensional gauge theory,
of the Hitchin systems, remains a mystery. 

This paper was inspired by a paper of Bershadsky, Johansen, Sadov and Vafa
~\cite{V}, about topologically reducing four-dimensional supersymmetric or
non-su\-per\-sym\-me\-tric gauge theories to two dimensions. 

In the 
${\cal N}=4$ four-dimensional case, it was found that the topologically
reduced two-dimensional $\sigma$-model had, as target space, a certain Hitchin
fibration of the cotangent bundle of unitary connections. 

In this paper, we show that the same is true for four-dimensional YM-
theories,
without any dimensional reduction, provided the entire infinite-dimensional
cotangent bundle is taken as target and the $\sigma$-model is
allowed to be four-dimensional and non-local. 

More precisely, we consider a YM theory on a four-torus, that we think as a
two-torus fibered on a two-torus, that is, as an elliptic manifold,
with gauge group $SU(N)$, without matter fields. 

We show that the partition function of this theory may, in a natural way, be
exactly represented as a non-linear and non-local four-dimensional
$\sigma$-model, whose target space is the cotangent space of two-dimensional
unitary connections, $T^* \cal {A}$, on the fiber torus fiberwise. 

The change of variables that leads to the cotangent bundle was suggested to 
the author by the existence of a well known duality transformation~\cite{Ha1,MM},
acting on the function field space of the YM functional integral.
This duality transformation maps the partition function of YM in four 
dimensions into the 
partition function of a theory that, expressed through the dual fields,
has an action with the same local part, up to some boundary terms, but
with the role of weak and strong coupling interchanged. 

In addition, the dual theory receives non-local contributions, unless certain
functional determinants, that appear in the course of
performing the duality transformation, cancel each other~\cite{MB}.
Though this cancellation will not be investigated in this paper, since the only
existence of the duality transformation is relevant to our main argument,
it was observed in  ref.~\cite{MB} that this cancellation of determinants would
be essentially equivalent to a property that in modern terms is called
$S$-duality for the partition function of the pure $SU(N)$ gauge theory with a
topological $\theta$-term. 

In the course of performing the duality transformation, in addition to the 
four-dimensional connection one-form $A$, the dual 
connection one-form $A^D$ and the dual field strength two-form $K$ are
introduced as auxiliary fields. Then, after integrating over the $A$ field and
some field redefinition, one gets the Bianchi identity constraint for the
$K$ field, that is solved as $K=F(A^D)$.
Now, our key point is that, instead
of eliminating all the $A$ and $K$ fields and keeping only the $A^D$ field, 
as in the usual definition of the duality transformation,
it is more interesting to 
integrate out all the fields but the ones that carry only the indices that
label tangent directions to the fiber torus of our 
elliptic fibration. We call this a partial or fiberwise duality transformation.
Employing complex coordinates on the fiber torus,
$(z,\bar {z})$, and on the base torus, $(u, 
\bar u)$, these fields are:
$\left(A_z,A_{\bar{z}},A^D_z,A^D_{\bar {z}}\right)$ and 
$K_{z \bar {z}}=F_{z \bar {z}}(A^D)
$. 

Remarkably, this integration can be done explicitly, by means of a trick,
integrating only Gaussian functionals.
The resulting functional integral is then a four-dimensional non-linear
and non-local $\sigma$-model, whose target space has coordinates
$\left(A_z,A_{\bar {z}},A^D_z,A^D_{\bar {z}}\right)$.
Performing the shift $A^D_z=A_z+\Psi_z,A^D_{\bar {z}}=A_{\bar {z}}+\Psi_{\bar 
{z}}$
with $\Psi$ a two-dimensional one-form fiberwise, the target space
of the $\sigma$-model becomes the cotangent space of the unitary
connections $T^* \cal{A}$ on the fiber torus fiberwise.

\section{A preparatory trick}

In this section we put the YM functional integral in a form suited for
performing the fiberwise duality transformation. Incidentally, we also 
write a gauge-fixed formula for the partition function in terms of only two 
physical polarizations.
The YM partition function is given by the formula:
\bea
Z&=&\int \exp\left[- \frac{1}{4g^2} \int \sum_{\mu \nu} 
Tr( F_{\mu \nu}^2) d^4x\right] DA \,,
\nonumber \\ 
F_{\mu \nu}&=&
\partial_{\mu}A_{\nu}-\partial_{\nu}A_{\mu}+i [A_{\mu},A_{\nu}] \, . 
\eea  
\pagebreak[3]

\noindent Tangent directions to the fiber torus are 
labelled by $\parallel$, while directions tangent to the base by $\perp$
. With this notation, introducing the auxiliary variable $E_{\perp}$, 
the partition function can be written as a Gaussian integral over 
$E_{\perp}$:
\bea
Z&=& \int \exp\left[- \frac{1}{2 g^2}\sum_{\parallel \perp}\int 
Tr \left(F_{\parallel}^2+ E_{\perp}^2-2i E_{\perp} F_{\perp}
+F_{\parallel \perp}^2\right) d^4x\right] DA DE_{\perp}\,.
\eea
The integration over the components of the connection transverse
to the fiber, $A_{\perp}$, is Gaussian and can be performed explicitly,
giving as a result the effective action for $E_{\perp}$ and 
$A_{\parallel}$, that is our desired preparatory result:
\bea
Z&=&\int \exp\left[- \frac{1}{2 g^2}\int Tr (F_{\parallel}^2+
	(\partial_{\perp}A_{\parallel})^2+ E_{\perp}^2) 
	d^4x\right] \times \nonumber \\
&& \hphantom{\int}\times \exp\left[ \frac{1}{2 g^2} \int Tr 
	\left[\left(\partial_{\bar{u}}(\partial_{\parallel}A_{\parallel}-E_{\perp})-j_{\bar{u}}\right)
	\left(-\Delta_{A_{\parallel}}-i \; ad_{E_{\perp}}\right)^{-1}\right.\right.\times\nonumber\\
&&\hphantom{\int}\hphantom{\times \exp~}
	\times\left(\partial_{u}
	(\partial_{\parallel}A_{\parallel}+E_{\perp})-j_{u}\right)\bigg]\Bigg]\times  
	\nonumber \\
&&\hphantom{\int}\times Det[-\Delta_{A_{\parallel}}-i \; ad_{E_{\perp}}]^{-1}
	DA_{\parallel} DE_{\perp} \,,
\label{2.3}\eea
where
\bea 
j_{u}=i [\partial_{u}A_{\parallel},A_{\parallel}]\,,\qquad
j_{\bar{u}}=i [\partial_{\bar{u}}A_{\parallel},A_{\parallel}]
\eea
and the sum over the appropriate space-time indices is understood.
$\Delta_{A_{\parallel}}$ is
the two-dimensional scalar Laplacian in the background of the connection 
$A_{\parallel}$
and $ad_{E_{\perp}}$ the adjoint action of the Lie algebra valued field 
$E_{\perp}$.
Incidentally $E_{\perp}$ can be integrated, with the help of a convenient 
choice of the 
gauge, that has also the advantage of eliminating some of the non-local terms
in the functional integral. The gauge choice~is:
\bea
\partial_{\parallel} A_{\parallel} + E_{\perp}=0 \, .
\eea
Inserting this gauge condition and the corresponding Faddeev-Popov 
determinant, we get:
\bea
Z&=&\int \exp\left[- \frac{1}{2 g^2}\int Tr \left(F_{\parallel}^2+ \left(
\partial_{\perp}A_{\parallel}\right)^
2+ \left(\partial_{\parallel} A_{\parallel}\right)^2\right)\right] \times 
\nonumber \\
&& \hphantom{\int}\times \exp\left[- \frac{1}{2 g^2} \int Tr  
 \left[(2\partial_{\bar{u}} \partial_{\parallel}A_{\parallel}-
j_{\bar{u}})
(-\Delta_{A_{\parallel}}+i \;  ad_{\partial_{\parallel} 
A_{\parallel}})^{-1}j_{u}\right]\right]
\times  \nonumber \\
&&\hphantom{\int}
\times Det\left[-\Delta_{A_{\parallel}}+i \, ad_{\partial_{\parallel} 
A_{\parallel}}\right]^{-
1} \Delta_{FP} DA_{\parallel}  \, .
\eea
The partition function is now a $\sigma$-model with only two polarizations.

\section{The duality transformation}

We remind the reader the well known four-dimensional duality transformation, 
that is the starting point of our argument, following the lines of 
refs.~\cite{Ha1,MM,MB}.
The Euclidean partition function of four-dimensional YM with a $\theta$-term
is defined by the formula:
\bea
Z&=&\int \exp\left[- \frac{1}{4g^2} \int \left(F^2 -i \tilde{\theta} 
F \tilde{F}\right)\right] DA \,,
\nonumber \\ 
\tilde{\theta}&=& \frac{g^2}{8\pi^2} \theta\,,  
\nonumber \\
F_{\mu \nu}&=&
\partial_{\mu}A_{\nu}-\partial_{\nu}A_{\mu}+i \left[A_{\mu},A_{\nu}\right] \, ,
\eea             
where the dual field strength is:
\bea
\tilde{F}_{\mu \nu}&=&
 \frac{1}{2}\epsilon_{\mu \nu \alpha \beta}  F_{\alpha \beta}
\eea 
and the trace and sum over the Euclidean indices are understood.
Employing the auxiliary variable $K$, the partition function can be 
written as a Gaussian integral over $K$:
\bea
Z&=&\int \exp\left[-\int \frac{g^2}{4\left(1+\tilde\theta^2\right)} 
\left(K^2+i \tilde{\theta} K
 \tilde K\right)+\frac{i}{2}\tilde{K} F(A)\right] DA DK \, .
\eea
After some manipulation, the following identity was found in ref.~\cite{MM}:
\bea
Z &=& \int \exp\left[-\int \frac{g^2}{4\left(1+\tilde{\theta}^2\right)}\left( K^2
+i \tilde{\theta} K \tilde{K}\right)+\frac{i}{2} \tilde{K} F(A^D) 
-\frac{i}{2} \xi_{\Sigma}\left(ad_{\tilde{K}}\right)^{-1} \xi_{\Sigma}\right]\times 
\nonumber \\
&& \hphantom{\int}
\times
\left|\Det (ad_{\tilde{K}})\right|^{\frac{1}{2}}\delta\left(d^*_{A^D} 
\tilde{K}\right) DA^D DK \, ,
\eea
where $\xi_{\Sigma}$ is a functional of $K$ that depends only on the boundary
values of $K$~\cite{MM}. The delta-functional constraint is the Bianchi
identity, that is solved by:
\bea
K&=&F(A^D)
\eea
up to some field-strength copies~\cite{Ha2, SS}.
In this way we get, for the final form of the partition function as a 
functional integral over the dual variables $A^D$~\cite{MB}:
\bea
Z &=& \int \exp\left[-\int
\frac{g^2}{4\left(1+\tilde{\theta}^2\right)}
\left( F(A^D)^2+i\tilde{\theta} \tilde{F}(A^D) F(A^D)\right)+\right. 
\nonumber \\
&& \hphantom{\int\exp \Big[}
 +
\left.\vphantom{\frac{g^2}{4\left(1+\tilde{\theta}^2\right)}} 
\frac{i}{2} \tilde{F}(A^D) F(A^D) 
-\frac{i}{2} \xi_{\Sigma}\left(ad_{\tilde{F}(A^D)}\right)^{-1} \xi_{\Sigma}\right] 
\times 
\nonumber \\
&& \hphantom{\int}
 \times
\left|\Det \left(ad_{\tilde{F}(A^D)}\right)\right|^{\frac{1}{2}} 
\left|\Det \left(d^*_{A^D}\right)\right|^{-1} DA^D \, . 
\eea
Instead of using as fundamental fields, through which to express
the QCD functional integral, the usual gauge potential, $A$, or its
dual, $A^D$, we choose $A_z,A_{\bar{z}},A^D_z,A^D_{\bar{z}}$.
Therefore we imagine of having integrated out all the other fields.
The main object that we want to compute is the corresponding effective 
action, $\Gamma$, in terms of these fields:
\bea
Z&=&\int \exp\left[-\Gamma\left(A_z,A_{\bar{z}},A^D_z,A^D_{\bar{z}}\right)\right]
DA_z DA_{\bar{z}} DA^D_z DA^D_{\bar{z}} \, .
\eea
Instead of computing it directly, which is rather cumbersome and perhaps
impossible using only Gaussian integrations, we make use of the following trick.
After the change of variables necessary to perform the duality 
transformation, the Bianchi identity constraint is implemented in the
functional integral via a delta-function, that is obtained integrating
over the original vector potential:
\bea
\delta(d^*_{A^D} \tilde{K})&=&\int
\exp\left[i \int\left(\partial_{\mu} \tilde{K}_{\mu \nu}-i\left[\tilde{K}_{\mu \nu}, 
A^D_{\mu}\right]\right) A_{\nu}\right] DA_{\nu} \, .
\eea
At this stage, inserting in the functional integral the constraint:
\bea
K_{z \bar{z}}=F_{z \bar{z}}(A^D)
\label{pippo}
\eea
is equivalent to multiply the partition function by an harmless
infinite factor, since eq.~(\ref{pippo}) is implied by the Bianchi identity constraint. 
If we want to introduce, instead, this constraint, before having performed
the change of variables needed to define $A^D$, we must
compensate the change of variables from $K_{z \bar{z}}$ to $A^D$ with a
Faddeev-Popov determinant, $\Delta_D$. Therefore we get:
\bea
Z&=&\int \exp\left[-\int \frac{g^2}{4\left(1+\tilde\theta^2\right)} 
\left(K^2+i \tilde{\theta} K
\tilde K\right)+\frac{i}{2}\tilde{K} F(A)\right] \times 
\nonumber \\
&&\hphantom{\int}
 \times  \delta \left(K_{z \bar{z}}-F_{z \bar{z}}(A^D)\right) \Delta_D 
DA^D_z DA^D_{\bar{z}} DA DK \, .
\label{3.10}\eea

\section{The fiberwise duality transformation}

We are now ready to change variables in order to embed
$T^* \cal {A}$ into the QCD functional integral. We simply set
\bea
E_{\perp}=\epsilon_{\perp \parallel} F_{\parallel}(A_D) \, 
\eea
into eq.~(\ref{2.3}), where $\epsilon_{\perp \parallel}$ is the rank
four normalized antisymmetric tensor and $A_D$ is a connection form on 
the fiber torus $T^2_f$ fiberwise.
We call this change of variables a fiberwise duality 
transformation because it has the structure of a duality transformation 
restricted to the fiber.
As in eq.~(\ref{3.10}), we introduce the resolution of the identity by means of the 
classical Faddeev-Popov trick:
\bea
1=\Delta_D(E)  \int \delta\left[E-F(A_D)\right] DA_D \, ,
\eea
where, in this section, $F(A_D)$ is the curvature two-form of the dual 
connection
one-form $A_D$ fiberwise. From now on we will refer to $A_{\parallel}$ as to
$A$. It is convenient to decompose $A_D$ into $A$ and an arbitrary one-form
$\Psi$:
\bea
A_D=A+\Psi \, .
\eea
Correspondingly, the Faddeev-Popov trick becomes, after shifting the $A_D$
integration~by~$A$:
\bea
1=\Delta_D(E,A)  \int \delta\left[E-F(A+\Psi)\right] D\Psi \, .
\eea
We may consider the fields $(A, \Psi)$ as the coordinates of $T^* \cal 
{A}$,
the cotangent space of unitary connections, $A$, on the fiber torus.
Of course $(A, \Psi)$ are four-dimensional fields as functions on space-time,
but they belong to $T^* \cal {A}$ fiberwise.
This gives, after adding the gauge-fixing condition, the desired embedding of  
$T^* {\cal {A}}/G$, fiberwise in the  functional integral:
\bea
Z&=&\int \exp\left[- \frac{N}{2\lambda} \int Tr\left( F_{A}^2+(d^* A)^2+
 (\partial_{\perp}A)^2\right) d^4x\right] \times 
\nonumber \\
&& \hphantom{\int}\times \exp\Bigg[\frac{N}{2\lambda}  \int Tr  
 \bigg[(\partial_{\bar{u}} (\partial_{\parallel}A_{\parallel}-F_{A+\Psi})-
j_{\bar{u}})(-\Delta_{A_{\parallel}}-i \; ad_{F_{A+\Psi}})^{-
1}\times\nonumber\\
&&\hphantom{\int\times \exp\Bigg[\frac{N}{2\lambda}  \int Tr~}\times
(\partial_{u} (\partial_{\parallel}A_{\parallel}+F_{A+\Psi})-j_{u})\bigg]\Bigg] \times  
\nonumber \\
&& \hphantom{\int}
\times \Det \left[-\Delta_A-i \, ad_{F_{A+\Psi}}\right]^{-1} \Delta_D(A,\Psi) \Delta_{FP} 
\times 
\nonumber \\[1ex]
&& \hphantom{\int}
\times \delta\left[d^*A+F_{A+\Psi}\right] DA D\Psi \, .
\eea

\end{document}